\documentclass[a4paper]{article}

\author{G. Gubbiotti\footnote{e-mail:
gubbiotti@mat.uniroma3.it},$\quad$  C. Scimiterna\footnote{e-mail:
scimiterna@fis.uniroma3.it}, $\quad$ D. Levi\footnote{e-mail:
decio.levi@roma3.infn.it}
\\
 Dipartimento di Matematica e Fisica, Universita' degli Studi Roma Tre,\\ e Sezione INFN di Roma Tre,\\ Via della Vasca Navale 84, 00146 Roma (Italy)
 }
\title{Linearizability and fake Lax pair for a consistent around the cube nonlinear non--autonomous quad--graph equation.}
\date{September 27$^{th}$, 2015}
\usepackage{rotating}

\usepackage{cite, amssymb}
\usepackage{braket}
\usepackage{amsmath}
\usepackage{booktabs}
\usepackage{amscd}
\usepackage[Q=yes]{examplep}
\usepackage{tikz,tikz-3dplot}
\usetikzlibrary{patterns}
\usetikzlibrary{arrows}
\usetikzlibrary{calc}
\usepackage{multirow}
\usepackage{rotating}
\usepackage{array}
\usepackage{longtable}
\usepackage{pdflscape}
\usepackage{collcell}
\usepackage{array}
\newcolumntype{C}{>{\displaystyle} c <{}}

\newcommand{\Z}{\mathbb{Z}}

\def\bea{\begin{eqnarray}}
\def\eea{\end{eqnarray}}
\def\ri{{\rm{i}}}

\newcommand{\Fp}[1]{\mathcal F^{(+)}_{#1}}
\newcommand{\Fm}[1]{\mathcal F^{(-)}_{#1}}

\newcounter{rmk}
\renewcommand{\thermk}{\arabic{rmk}}
{\refstepcounter{rmk}\vspace{6pt}\noindent\ignorespaces\textbf{Remark
\thermk:}}{\vspace{6pt}\par}

\renewcommand{\epsilon}{\varepsilon}

\allowdisplaybreaks

\begin{document}

\maketitle
\begin{abstract}
We discuss the linearization of a non--autonomous nonlinear partial difference equation belonging to the Boll classification of quad--graph equations consistent around the cube.  We show that its Lax pair is fake. We present its generalized symmetries  which turn out to be  non--autonomous and depending  on an arbitrary function of the dependent variables defined in two lattice points. These generalized symmetries are differential difference equations which, in some case,  admit peculiar B\"acklund transformations. 
\end{abstract}
%%%%%%%%%%%%%%%%%%%%%
%%%%%%%%%%%%%%%%%%%%%
\section{Introduction.}
%%%%%%%%%%%%%%%%%%%%%
%%%%%%%%%%%%%%%%%%%%%
Symmetries and commuting flows have been in a way or in an'other at the base of integrability. B\"acklund transformations and nonlinear superposition rules \cite{lb,ns} paved the way to the discretization of integrable systems. 

A first version of the integrability criteria  denoted Consistency Around the Cube ($CAC$) can be found in the work of Doliwa and Santini \cite{ds}.

In recent years $CAC$ has been a source of many results in the classification of nonlinear difference equations. Its importance relays in the fact that provides B\"acklund transforms \cite{BoS,BrH,N,NW} and as a consequence the existence of a zero curvature representation or Lax pairs, which, as it is well known \cite{yamilov2006}   are associated to both linearizable and integrable equations.  %While, as far as we know  the existence of conserved quantities of all orders  is a property characteristic only of  integrable equations.

 The first attempt to carry out a classification of partial difference equations using  the CAC condition  has been presented in \cite{ABS03} assuming that the equations on all faces of the cube  were the same  form. The result is a class of discrete equations  formed by systems living on quad-graphs, whose basic building blocks are equations on quadrilaterals of the type 
 \bea \label{1.1}
 A\left(x,x_{1},x_{2},x_{12};\alpha_{1},\alpha_{2}\right)=0,
 \eea
  where the four fields $x$, $x_{1}$, $x_{2}$ and $x_{12}\in{\mathcal C}$  are assigned to the four vertexes of a quadrilateral (later in (\ref{1.4}) denoted as $x_1$, $x_2$, $x_4$ and $x_4$ or equivalently on a lattice of indices $m$ and $n$ as $x_{m,n}$, $x_{m+1,n}$, $x_{m,n+1}$ and $x_{m+1,n+1}$) and the parameters $\alpha_{i}\in{\mathcal C}$, $i=1$, $2$ to its edges ( $\alpha_2=\alpha_2(n)$, $\alpha_1=\alpha_1(m)$).
 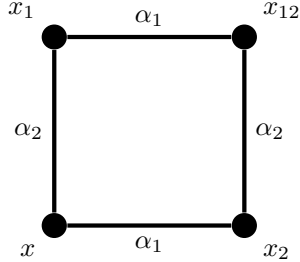
\begin{figure}[htbp]
   \centering
   \begin{tikzpicture}[auto]
      \node (x1) at (0,0) [circle,fill,label=-135:$x$] {};
      \node (x4) at (0,2.5) [circle,fill,label=135:$x_{1}$] {};
      \node (x2) at (2.5,0) [circle,fill,label=-45:$x_{2}$] {};
      \node (x3) at (2.5,2.5) [circle,fill,label=45:$x_{12}$] {};
      \draw [ultra thick] (x2) to node {$\alpha_{1}$} (x1);
      \draw [ultra thick] (x4) to node {$\alpha_{1}$} (x3);
      \draw [ultra thick] (x3) to node {$\alpha_{2}$} (x2);
      \draw [ultra thick] (x1) to node {$\alpha_{2}$} (x4);
   %   \draw [ultra thick] (x2) to (x4);
      %\draw [ultra thick] (x1) to (x3);
   \end{tikzpicture} 
   \caption{Quad-graph }
\label{fig:bp}
\end{figure}
 In the notation defined above, if the quadrilateral is constructed designing  two independent directions $1$, $2$ starting from an origin vertex on which the field $x$ is assigned, then the subscript $1$, $2$ provide the field in the vertex shifted by $\alpha_1$, $\alpha_2$ along the direction $1$, $2$ from the origin, while $12$ refers to the field in the remaining vertex of the quadrilateral. Moreover the function $A\left(x,x_{1},x_{2},x_{12};\alpha_{1},\alpha_{2}\right)$ is assumed to be affine linear in each argument (multilinearity) with coefficients depending on the two edge parameters and invariant under the discrete group $D_{4}$ of square symmetries
\bea
\nonumber A\left(x,x_{1},x_{2},x_{12};\alpha_{1},\alpha_{2}\right)&=&\epsilon A\left(x,x_{2},x_{1},x_{12};\alpha_{2},\alpha_{1}\right)\\ \nonumber &=&\sigma A\left(x_{1},x,x_{12},x_{2};\alpha_{1},\alpha_{2}\right),\qquad  \epsilon,\sigma=\pm 1.
\eea
 \noindent A last simplifying hypothesis is the so called \emph{tetrahedron property} 
 \begin{figure}[htbp] 
   \centering
   \begin{tikzpicture}[auto,scale=0.8]
      \node (x) at (0,0) [circle,fill,label=-45:$x$] {};
      \node (x1) at (4,0) [circle,fill,label=-45:$x_{1}$] {};
      \node (x2) at (1.5,1.5) [circle,fill,label=-45:$x_{2}$] {};
      \node (x3) at (0,4) [circle,fill,label=-45:$x_{3}$] {};
      \node (x12) at (5.5,1.5) [circle,fill,label=-45:$x_{12}$] {};
      \node (x13) at (4,4) [circle,fill,label=-45:$x_{13}$] {};
      \node (x23) at (1.5,5.5) [circle,fill,label=-45:$x_{23}$] {};
      \node (x123) at (5.5,5.5) [circle,fill,label=-45:$x_{123}$] {};
      \node (A) at (2.75,0.75) {$A$};
      \node (Aq) at (2.75,4.75) {$\bar A$};
      \node (B) at (0.75,2.75) {$B$};
      \node (Bq) at (4.75,2.75) {$\bar B$};
      \node (C) at (2,2) {$C$};
      \node (Cq) at (3.5,3.5) {$\bar C$};
      \draw (x) to (x1) to (x12) to (x123) to (x23) to (x3) to (x);
      \draw (x3) to (x13) to (x1);
      \draw (x13) to (x123);
      \draw [dashed] (x) to (x2) to (x12);
      \draw [dashed] (x2) to (x23);
      \draw [dotted,thick] (A) to (Aq);
      \draw [dotted,thick] (B) to (Bq);
      \draw [dotted,thick] (C) to (Cq);
   \end{tikzpicture} 
   \caption{Equations on a Cube}
   \label{fig:cube}
\end{figure}
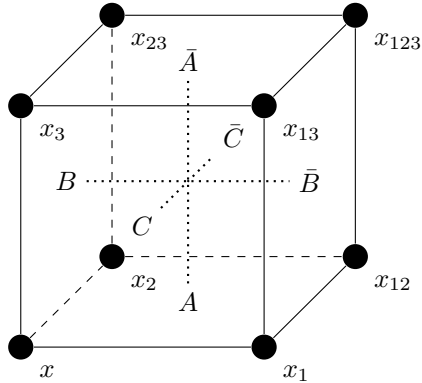
which amounts to require that, starting from arbitrary initial values $x$, $x_{1}$, $x_{2}$ and $x_{3}$,  the function $x_{123}\doteq x_{123}\left(x,x_{1},x_{2},x_{3};\alpha_{1},\alpha_{2},\alpha_{3}\right)$  does not depend on $x$. On a 3--D lattice of indices $m$, $n$ and $p$, on the cube depicted in Fig. \ref{fig:cube}, $x=x_{m,n,p}$, $x_1=x_{m+1,n,p}$, $x_2=x_{m,n+1,p}$ and $x_3=x_{m,n,p+1}$. Then $x_{123}=x_{m+1,n+1,p+1}$. The lattice equation $A(x_{m,n,p}, x_{m+1,n,p}, x_{m,n+1,p}, x_{m+1,n+1,p}; \alpha_1(m), \alpha_2(n))=0$ is defined on the square of indices $m$ and $n$ and depends parametrically on the index $p$ while $\bar A=0$ is the same as $A=0$ in the indices $m$ and $n$ but for a different value of $p$,  $p+1$. The equation $B=0$ is defined on the square of indices $m$ and $p$ and depends parametrically on the index $n$ while $\bar B=0$ is defined on the square of indices $n$ and $p$ and depends parametrically on the index $m$. The final result of the classification is given by a set of two lists of equations, $H$ and $Q$, for a total of seven consistent systems (up to a common M\"obius transformation of the field variables and point transformations of edge parameters) are presented in \cite{ABS03}.
So, in Fig. \ref{fig:cube}, on a 2--D lattice, $\bar A$ represents a copy of the equation $A$ (one of the seven equations $H$ and $Q$) and on the faces $B$ and $\bar B$ we have a relation between a solution of $A$ and one of $\bar A$, i.e. an auto--B\"acklund transformation for $A$. By going over to projective space the auto--B\"acklund transformation provide the Lax pair.

 The same authors in \cite{ABS09} considered a more general perspective in the classification problem. They assumed that the faces of the consistency cube $A$, $B$, $C$ and $\bar A$, $\bar B$ and $\bar C$ could carry a priori different quad-equations without assuming either the $D_{4}$ symmetry or the tetrahedron property. 
 They considered six-tuples of (a priori different) quad-equations assigned to the faces of a 3D cube:
\begin{align}\label{system}
&A\left(x,x_{1},x_{2},x_{12}; \alpha_1, \alpha_2\right)=0,&
&\bar{A}\left(x_{3},x_{13},x_{23},x_{123}; \alpha_1, \alpha_2\right)=0,\notag\\
&B\left(x,x_{2},x_{3},x_{23}, \alpha_2, \alpha_3\right)=0,&
&\bar{B}\left(x_{1},x_{12},x_{13},x_{123}, \alpha_2, \alpha_3\right)=0,\\
&C\left(x,x_{1},x_{3},x_{13}; \alpha_1, \alpha_3\right)=0,&
&\bar{C}\left(x_{2},x_{12},x_{23},x_{123}; \alpha_1, \alpha_3\right)=0,\notag
\end{align}
see Fig. \ref{fig:cube}. Such a six-tuple is \emph{3D consistent} if, for arbitrary initial data $x$, $x_{1}$, $x_{2}$ and $x_{3}$,
the three values for $x_{123}$ (calculated by using $\bar{A}=0$, $\bar{B}=0$ or $\bar{C}=0$) coincide.
  As a result in \cite{ABS09} they reobtained  the $Q-$type equations of \cite{ABS03} and showed some new examples of quad-equations of type $H$ which turn out to be deformations of the previously one obtained \cite{ABS03}).
 
In  \cite{Boll11,Boll12a,Boll12b}, Boll classified all the consistent quad-equations possessing the tetrahedron property without any other additional assumption. All the results were summarized in a set of theorems,  from Theorem {\bf 3.9} to Theorem {\bf 3.14} in \cite{Boll12b}, listing all the consistent six-tuples configurations up to $\left(\right.$M\"ob$\left.\right)^8$, the group of independent M\"obius transformations of the eight fields on the vertexes of the consistency cube. Defining for each equation (\ref{1.1})  the accompanying biquadratics
\bea \label{1.4}
A^{i,j} \equiv A^{i,j}(x_i,x_j)=A_{,x_m}A_{,x_n}-A A_{,x_m x_n},
\eea
where $\{m,n\}$ is the complement of $\{i,j\}$ in $\{1,2,3,4\}$, all the quad-equations fall into three disjoint families: $Q-$type (no degenerate biquadratic), $H^{4}-$type (four biquadratics are degenerate) and $H^{6}-$type (all of the six biquadratics are degenerate).

\begin{figure}[htpb]
\centering
\begin{tikzpicture}[scale=2.5]
    \draw [pattern=north west lines,thick] (0,0) rectangle (1,1);
    \draw [pattern=north east lines,thick] (1,1) rectangle (2,2);
    \draw [pattern=vertical lines,thick] (1,0) rectangle (2,1);
    \draw [pattern=horizontal lines,thick] (0,1) rectangle (1,2);
    \foreach \x in {0,...,2}{% Two indices running over each
        \foreach \y in {0,...,2}{% node on the grid we have drawn 
            \node[draw,circle,inner sep=2pt,fill] at (\x,\y) {};
        }
    }
    \node[below left] at (0,0) {$x$};
    \node[below] at (1,0) {$x_1$};
    \node[below right] at (2,0) {$x$};
    \node[left] at (0,1) {$x_2$};
    \node[below left] at (1,1) {$x_{12}$};
    \node[right] at (2,1) {$x_2$};
    \node[above left] at (0,2) {$x$};
    \node[above] at (1,2) {$x_1$};
    \node[above right] at (2,2) {$x$};
    \node[] at (1/2,1/2) {$Q$};
    \node[] at (1/2+1,1/2) {$|Q$};
    \node[] at (1/2,1/2+1) {$\underline{Q}$};
    \node[] at (1/2+1,1/2+1) {$|\underline{Q}$};
\end{tikzpicture}
\caption{The ``four colors'' lattice}
\label{fig:elcell}
\end{figure}
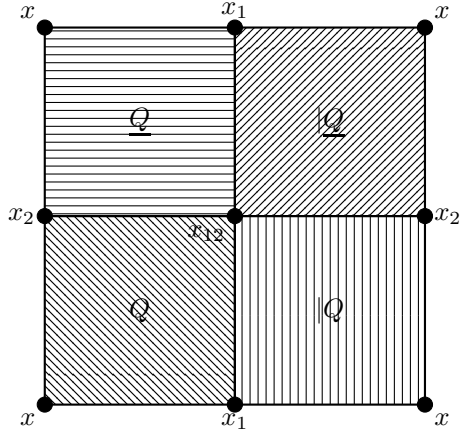

\indent It's worth emphasizing that all the classification results holds locally, in the sense that everything is stated on a single quadrilateral cell or on a single cube. The non secondary problem of the embedding in a $2D$/$3D$ lattice of the single cell/single cube equations, so as to preserve $3D$ consistency, was already discussed in \cite{ABS09} introducing the concept of Black--White (BW) lattice. 
One way to solve this problem, is to  embed  (\ref{system}) into a $\Z^{2}$ lattice with an elementary cell
of dimension greater than one.  In such a case  the equation $Q=Q(x,x_{1},x_{2},x_{12}; \alpha_1, \alpha_2)$ can be extended to a lattice and the lattice equation will become integrable or linearizable. 
To do so, following \cite{Boll11}, we reflect the square with respect 
to the normal to its right and top sides and then complete
a $2\times2$ lattice by reflecting again one of the obtained equation with respect to the
other direction\footnote{Let us note that, whatsoever side we reflect,  the result of the last reflection is the same.}. 
Such procedure is graphically described in Figure \ref{fig:elcell}, and at the
level of the quad equation this correspond to construct the three  equations obtained from
$Q=Q(x,x_{1},x_{2},x_{12}; \alpha_1, \alpha_2)$ by flipping its fields:
\begin{subequations}
\begin{align}
Q &=Q(x,x_{1},x_{2},x_{12},\alpha_{1},\alpha_{2}) =0,
\\
|Q &=Q(x_{1},x,x_{12},x_{2},\alpha_{1},\alpha_{2}) =0,
\\
\underline{Q} &= Q(x_{2},x_{12},x,x_{1},\alpha_{1},\alpha_{2}) =0,
\\
|\underline{Q} &= Q(x_{12},x_{2},x_{1},x,\alpha_{1},\alpha_{2}) =0.
\end{align}
\label{eqn:dysys1}
\end{subequations}
By paving the whole $\Z^{2}$ with such equation we will get a partial difference equation,
which we can in principle study with the known methods. Since
\emph{a priori}  $Q\neq |Q \neq \underline{Q} \neq |\underline{Q}$  the obtained lattice will be
a four color lattice, i.e. an extension of the BW lattice
 \cite{HV,XP}.

 If the equation $Q=0$ is invariant under the action of $D_{4}$
we have that:
\begin{equation}
    Q = |Q = \underbar{Q} = |\underbar{Q}
    \label{eq:symmsquare}
\end{equation}
implying that the elementary cell is actually of dimension one,
and we fall into the case of the ABS classification.
When 
\begin{equation}
    \begin{aligned}
    Q(x,x_{1},x_{2},x_{12},\alpha_{1},\alpha_{2}) &=
    \sigma Q(x,x_{2},x_{1},x_{12},\alpha_{2},\alpha_{1})
    \\
    &=\varepsilon Q(x_{12},x_{1},x_{2},x,\alpha_{2},\alpha_{1}),
    \end{aligned}
    \qquad
    \sigma,\varepsilon\in\Set{\pm1}
    \label{eq:rhombsymm}
\end{equation}
i.e.
\bea \label{rhombic}
Q=|\underline{Q}, \qquad \underline{Q}=|Q
\eea
 the equation has {\it rhombic symmetry} while if \begin{equation}
    Q(x,x_{1},x_{2},x_{12},\alpha_{1},\alpha_{2}) = Q(x_{1},x,x_{12},x_{2},\alpha_{1},\alpha_{2}).
    \label{eq:trapsymm}
\end{equation}
i.e.  
 \bea \label{trapezoidal1}
 Q=|Q, \qquad \underline{Q}=|\underline{Q}
 \eea
  or 
  \bea \label{trapezoidal2}
  Q=\underline{Q},\qquad |Q=|\underline{Q}
  \eea
   we say that the equation has {\it trapezoidal symmetry}.
A detailed study of all the lattices derived from the the \emph{rhombic} $H^{4}$ family, including the construction of the  Lax pairs, B$\ddot{a}$cklund transformations and infinite hierarchies of generalized symmetries, was presented in \cite{XP}. 
In \cite{GLS2015} they studied all the equations of the Boll classification not already considered in the previous literature \cite{ABS03,ABS09,XP} and showed by using the algebraic entropy \cite{Tremblay2001,Viallet2006} that they are linearizable and provided their explicit linearization.

A general procedure for the  embedding is  given in \cite{Boll11}, \cite{Boll12b}. Different embeddings in $3D$ consistent lattices resulting either in integrable or non integrable equations are discussed in \cite{HV} using an algebraic entropy analysis.

 After the original work of Adler Bobenko and Suris there have been various attempts to  simplify the requirements imposed on consistent quad-equations. In this way four non tetrahedral models, three of them with $D_{4}$ symmetry, were presented in \cite{H04,H05}. All of these  models are linear equations or (more or less trivially) linearizable \cite{RJ}. Other non tetrahedral, consistent systems of linear quadrilateral lattice equations were studied in \cite{ABS09,At09}.

The existence of fake Lax pairs is well known but the phenomenon is not widely understood.
The term Lax pair refers to  a pair of linear equations (an overdetermined system)  associated with a nonlinear integrable system through a compatibility condition. The most important property of a Lax pair is that it should prove the integrability of the associated nonlinear system and provide information about nontrivial solutions to nonlinear integrable system. 

Fake Lax pairs are Lax pairs which tell us  nothing about the integrability of the associated nonlinear system. Fake Lax pairs often appear very similar to their integrable counterparts and experts in the area of integrable systems continue to inadvertently publish fake Lax pairs that they believe are real (see \cite{hb} for examples). %As such, there is serious need for a straightforward method to distinguish between real and fake Lax pairs.
Moreover fake Lax pairs are usually associated to linearizable systems and to prove the fakeness of a Lax pair is a way to prove the linearizability of a system.

%In this letter, we provide two very simple methods to identify fake Lax pairs. 
Lax pairs can be discrete or continuous, matrix or scalar, used for inverse scattering or isomonodromy, and fake Lax pairs reside in all of these categories. Although there are many references to fake Lax pairs in the literature, the most famous being \cite{cn91}, there are fewer articles that set out to explain what fake Lax pairs are and how to identify them. Various methods have been given to identify fake Lax pairs \cite{hb, lsz90, lll10, m10, m04, s01, s02}.
%The methods to identify fake Lax pairs presented in this article is applicable to any kind of Lax pair that contains a non-removable spectral variable. The question of inserting a non-removable spectral variable into a Lax pair that does not have one is explored in \cite{m10}.

%The methods outlined in this article highlight properties that are sufficient for a Lax pair to be fake, but are not necessary. It is possible that a Lax pair could pass these tests and still be fake, however, every fake Lax pair known to the authors fails both tests.

 %have been limited in their applicability and often difficult to apply. In contrast, the tests given here are easily comprehensible and widely applicable.

 In the following in Section 2 we focus on the simplest of the $H^4$ equations,  the ${}_{t}H_{1}^\epsilon$ equation, and present its Lax pair. In Section 3 we show its direct linearizability and that the obtained Lax pair is effectively fake.  In Section 4 we present its generalized symmetries which turn out to be differential difference equations which have, by construction B\"acklund transformations. Section 5 is devoted to some final conclusions.  

%%%%%%%%%%%%%%%%%%%%%%%%%
%%%%%%%%%%%%%%%%%%%%%%%%%
\section{The ${}_{t}H_{1}^\epsilon$ equation.}
%%%%%%%%%%%%%%%%%%%%%%%%%
%%%%%%%%%%%%%%%%%%%%%%%%%
Before defining the equation ${}_{t}H_{1}^\epsilon$ we need to introduce some notation:
\bea
&&\nonumber{\mathcal F}_{p}^{\left(\pm\right)}\doteq\frac{1\pm\left(-1\right)^p}{2},\ \ \ \alpha_{1}\doteq\alpha_{1}\left(m\right),\ \ \ \alpha_{2}\doteq\alpha_{2}\left(n\right),\ \ \ \alpha_{3}\doteq\alpha_{3}\left(p\right),\\
\eea
where $\alpha_1$, $\alpha_2$  are  constants which appear in the equation and might depend on the lattice variable and play the role of the lattice spacing when carrying out the continuous limit. $\alpha_3$ will appear in the B\"acklund transformations and will play the role of the spectral paramether.
\bea
&&\nonumber\psi_{m,n+1}= \ell_{m,n}\cdot L_{m,n}\psi_{m,n},\ \ \ \psi_{m+1,n}={\mbox{m}}_{m,n}\cdot M_{m,n}\psi_{m,n},\ \ \ \tau_{m,n}\doteq\frac{{\mbox{m}}_{m,n} \ell_{m+1,n}}{ \ell_{m,n}{\mbox{m}}_{m,n+1}},\\ \label{laxr} && \qquad \qquad \qquad  \tau_{m,n}\cdot\ L_{m+1,n}\cdot M_{m,n}- M_{m,n+1}\cdot L_{m,n}=0,\\
&&\nonumber \ell_{m,n}\doteq \ell\left(m,n,x_{m,n},x_{m,n+1};\alpha_{3},\alpha_{1},\epsilon\right),\ \ \  L_{m,n}\doteq L\left(m,n,x_{m,n},x_{m,n+1};\alpha_{3},\alpha_{1},\epsilon\right),\\
&&\nonumber {\mbox{ m}}_{m,n}\doteq{\mbox{m}}\left(m,n,x_{m,n},x_{m+1,n};\alpha_{2},\alpha_{3},\epsilon\right),\ \ \  M_{m,n}\doteq M\left(m,n,x_{m,n},x_{m+1,n};\alpha_{2},\alpha_{3},\epsilon\right).
\eea
 The  elements $ L$ and $ M$  
represent the Lax pair on the $m$ and $n$ lattice and (\ref{laxr}) is the corresponding Lax equation on the lattice. The related normalization coefficients are denoted  $ \ell$ and ${\mbox{m}}$ and $\tau$ is their ratio which appear in the Lax equation. $\alpha_3$ place the role of the spectral parameter.

 Then  the ${}_{t}H_{1}^\epsilon$ equation reads:
\bea \label{eql}
\left(x_{m,n}-x_{m+1,n}\right)\left(x_{m,n+1}-x_{m+1,n+1}\right)-\epsilon^2\alpha_{2}\left({\mathcal F}_{n}^{\left(+\right)}x_{m,n+1}x_{m+1,n+1}+{\mathcal F}_{n}^{\left(-\right)}x_{m,n}x_{m+1,n}\right)-\alpha_{2}=0.
\eea
(\ref{eql}) it is obtained from the cell equation 
\bea \label{eqc}
\left(x-x_{2}\right)\left(x_{3}-x_{23}\right)-\alpha_{2}\left(1+\epsilon^2x_{3}x_{23}\right)=0,
\eea
by the procedure schematized in the introduction and described in detail in \cite{GLS2015}.

Its Lax pair and normalization coefficients are:
\bea \label{co-lax}
&& L_{m,n}=\left(\begin{array}{cc}
x_{m,n+1} & -x_{m,n}x_{m,n+1}+\alpha_{3}\\
1 & -x_{m,n}\end{array}\right)-\epsilon^2\alpha_{3}\left(\begin{array}{cc}
-{\mathcal F}_{n}^{\left(-\right)}x_{m,n} & 0\\
0 & {\mathcal F}_{n}^{\left(+\right)}x_{m,n+1}\end{array}\right),\label{Lax1}\\ \label{Lax2}
&& M_{m,n}=\left(\begin{array}{cc}
\alpha_{3}\left(x_{m,n}-x_{m+1,n}\right)+\alpha_{2}x_{m+1,n} & -\alpha_{2}x_{m,n}x_{m+1,n}\\ 
\alpha_{2} & \alpha_{1}\left(x_{m,n}-x_{m+1,n}\right)-\alpha_{2}x_{m,n}\end{array}\right)-\\ &&\nonumber  \quad - \epsilon^2\alpha_{3}\alpha_{2}\left(\alpha_{3}-\alpha_{2}\right){\mathcal F}_{n}^{\left(+\right)}\left(\begin{array}{cc}
0 & 1\\
0 & 0\end{array}\right), \qquad
\tau_{m,n}=\frac{\alpha_{2}\left(1+\epsilon^2{\mathcal F}_{n}^{\left(-\right)}x_{m,n}^2+\epsilon^2{\mathcal F}_{n}^{\left(+\right)}x_{m,n+1}^2\right)}{\left(x_{m,n}-x_{m+1,n}\right)^2+\epsilon^2\alpha_{2}^2{\mathcal F}_{n}^{\left(+\right)}},\\
%&&\nonumber \ell_{m,n}=\frac{1}{\sqrt{1+\epsilon^2\left({\mathcal F}_{n}^{\left(-\right)}x_{m,n}^2+{\mathcal F}_{n}^{\left(+\right)}x_{m,n+1}^2\right)}},\ \ \ {\mbox{m}}_{m,n}=\frac{1}{\sqrt{\left(x_{m,n}-x_{m+1,n}\right)^2+\epsilon^2\alpha_{2}^2{\mathcal F}_{n}^{\left(+\right)}}},\\
&& \ell_{m,n}=\frac{1}{1\mp\ri\epsilon\left({\mathcal F}_{n}^{\left(-\right)}x_{m,n}-{\mathcal F}_{n}^{\left(+\right)}x_{m,n+1}\right)},\ \ \ {\mbox{m}}=\frac{1}{x_{m,n}-x_{m+1,n}\pm\ri\epsilon\alpha_{2}{\mathcal H}_{n}^{\left(+\right)}},\label{Lax3}
\eea
where, with a completely non trivial calculation, one has been able to rewrite the connection formulas  (\ref{Lax3}) without using the standard square root terms. 
The equations associated to the other sides of the cube are:
\bea \label{bt1}
&&\alpha_{2}\left(x_{m,p}-x_{m,p+1}\right)\left(x_{m+1,p}-x_{m+1,p+1}\right)-\alpha_{3}\left(x_{m,p}-x_{m+1,p}\right)\left(x_{m,p+1}-x_{m+1,p+1}\right)+\\
\nonumber&& \qquad \qquad \qquad +\epsilon^2\alpha_{3}\alpha_{2}\left(\alpha_{3}-\alpha_{2}\right){\mathcal F}_{q}^{\left(+\right)}=0,\\
 \nonumber
&&\left(x_{n,q}-x_{n,q+1}\right)\left(x_{n+1,q}-x_{n+1,q+1}\right)-\epsilon^2\alpha_{3}\left({\mathcal F}_{n}^{\left(+\right)}x_{n+1,q}x_{n+1,q+1}+{\mathcal F}_{n}^{\left(-\right)}x_{n,q}x_{n,q+1}\right)-\alpha_{3}=0.
\eea
Eqs. (\ref{bt1}) are the B\"acklund transformations for the ${}_{t}H_{1}^\epsilon$ equation when the lattice variables $p+1$ and $q+1$ are interpreted as indices indicating a new solution. By going over to projective space (\ref{bt1}) provide the Lax pair (\ref{Lax2}).
 
%%%%%%%%%%%%%%%%%%%%%%%%%
%%%%%%%%%%%%%%%%%%%%%%%%%
\section{Linearization and fake Lax pairs.}
%%%%%%%%%%%%%%%%%%%%%%%%%
%%%%%%%%%%%%%%%%%%%%%%%%%
Let us now analyze equation (\ref{eql}). 
 In (\ref{eql}) it is always possible to suppose $\alpha_{2}\not=0$, otherwise the equation degenerates into $\left(x_{m,n}-x_{m+1,n}\right)\left(x_{m,n+1}-x_{m+1,n+1}\right)=0$ whose solution is trivial on the lattice. Let us define $x_{m,2k}\doteq w_{m,k}$, $x_{m,2k+1}\doteq z_{m,k}$; then we have the following system of two coupled autonomous difference equations
\begin{subequations}
\bea
\left(w_{m,k}-w_{m+1,k}\right)\left(z_{m,k}-z_{m+1,k}\right)-\epsilon^2\alpha_{2}z_{m,k}z_{m+1,k}-\alpha_{2}=0,\label{Krur1}\\
\left(w_{m,k+1}-w_{m+1,k+1}\right)\left(z_{m,k}-z_{m+1,k}\right)-\epsilon^2\alpha_{2}z_{m,k}z_{m+1,k}-\alpha_{2}=0.\label{Krur2}
\eea
\end{subequations}\\
\noindent Subtracting (\ref{Krur2}) to (\ref{Krur1}), we obtain
\bea \label{20}
\left(w_{m,k}-w_{m+1,k}-w_{m,k+1}+w_{m+1,k+1}\right)\left(z_{m,k}-z_{m+1,k}\right)=0.
\eea\\
\noindent At this point the solution of the system bifurcates:\\
\begin{itemize}
\item {\bf Case 1}: if $z_{m,k}=f_{k}$, where $f_k$ is a generic function of its argument, equation (\ref{20}) is satisfied and  from (\ref{Krur1}) or (\ref{Krur2}) we have that $\epsilon\not=0$ and, solving for $f_{k}$, one gets
\bea \label{21}
f_{k}=\pm\frac{\ri}{\epsilon}.
\eea
\item {\bf Case 2}: if $z_{m,k}\not=f_k$, with $f_k$ given in (\ref{21}),  one has $w_{m,k}=g_{m}+h_{k}$, where $g_m$ and $h_k$ are arbitrary functions of their argument. Hence (\ref{Krur2}) and (\ref{Krur1}) reduce to
\bea
\epsilon^2 z_{m,k}z_{m+1,k}+\kappa_{m}\left(z_{m,k}-z_{m+1,k}\right)+1=0,\ \ \ \kappa_m\doteq\frac{g_{m+1}-g_{m}}{\alpha_{2}},\label{Krur3}
\eea\\
\noindent so that two sub-cases emerge:
\begin{itemize}
\item {\bf Sub-case 2.1}: if $\epsilon=0$, (\ref{Krur3}) implies $\kappa_{m}\not=0$, so that, solving,
\bea
z_{m+1,k}-z_{m,k}=\frac{1}{\kappa_{m}},
\eea\\
we get
\begin{subequations}
\bea
z_{m,k}=j_{k}+\sum_{l=m_{0}}^{m-1}\frac{1}{\kappa_{l}},\ \ \ m\geq m_{0}+1,\\
z_{m,k}=j_{k}-\sum_{l=m}^{m_{0}-1}\frac{1}{\kappa_{l}},\ \ \ m\leq m_{0}-1,
\eea
\end{subequations}\\
\noindent where $j_k=z_{m_{0},k}$ is a generic integration function of its argument.
\item {\bf Sub-case 2.2}: if $\epsilon\not=0$, (\ref{Krur3}) is a discrete Riccati equation which can be linearized by the M$\ddot{o}$bius transformation $z_{m,k}\doteq\frac{\ri}{\epsilon}\frac{y_{m,k}-1}{y_{m,k}+1}$ to
\bea \label{25}
\left(\ri \kappa_{m}-\epsilon\right)y_{m+1,k}=\left(\ri \kappa_{m}+\epsilon\right)y_{m,k},
\eea\\
\noindent which, as $\kappa_{m}\not=\pm\ri\epsilon$ because otherwise $y_{m,k}=0$ and $z_{m,k}=-\ri/\epsilon$.  Eq. (\ref{25}) implies
\begin{subequations}
\bea
y_{m,k}=j_{k}\prod_{l=m_{0}}^{m-1}\frac{\ri \kappa_{l}+\epsilon}{\ri \kappa_{l}-\epsilon},\ \ \ m\geq m_{0}+1,\\
y_{m,k}=j_{k}\prod_{l=m}^{m_{0}-1}\frac{\ri \kappa_{l}-\epsilon}{\ri \kappa_{l}+\epsilon},\ \ \ m\leq m_{0}-1,
\eea
\end{subequations}\\
\noindent where $j_k=y_{m_{0},k}$ is another arbitrary integration function of its argument.
\end{itemize}
\end{itemize}
\noindent In conclusion we have always completely integrated the original system.\\

Let us note that in the case $\epsilon=0$ when \eqref{eql}
        becomes
        \begin{equation}
            \left( x_{m,n}-x_{m+1,n} \right)\left( x_{m,n+1}-x_{m+1,n+1} \right)-\alpha_{2}=0,
            \label{eq:tH1e0}
        \end{equation}
 the contact M\"obius-type
        transformation
        \begin{equation}
            x_{m,n} - x_{m+1,n} = \sqrt{\alpha_{2}}\frac{2w_{m,n}+1-\alpha_{2}}{%
                2 w_{m,n}+1+\alpha_{2}},
            \label{eq:mobh1_0lin}
        \end{equation}
        brings (\ref{eql}) into the following first order linear equation:
        \begin{equation}
            w_{m,n+1}+w_{m,n} +1  = 0,
            \label{eq:h1_0linmob}
        \end{equation}
        whose  solution is:
        \begin{equation}
            x_{m,n} = x_{0,n} -\sqrt{\alpha_{2}}\sum_{l=1}^{m}
            \frac{2 w_{l,0}\left( -1 \right)^{n}+\left( -1 \right)^{n}-\alpha_{2}}{%
                2 w_{l,0}\left( -1 \right)^{n}+\left( -1 \right)^{n}+\alpha_{2}}.
            \label{eq:th1e0}
        \end{equation}
        Here $x_{0,n}$ and $w_{m,0}$ are two arbitrary integration functions. 
        
An'other linearizing transformation is given by
\begin{equation}
            x_{m,n} - x_{m+1,n} = \sqrt{\alpha_{2}} e^z_{m,n},
            \label{31}
\end{equation}
bringing (\ref{eql}) into the following first order linear equation:
\bea \label{32}
        z_{m,n+1}+z_{m,n}= 2*i \pi \kappa,
        \eea 
where $\kappa$ is an arbitrary entire paramether.

 The linearization can also be achieved using the Lax pair (\ref{Lax1}-\ref{Lax3}). Introducing the fields $\rho_{m,k}\doteq\tilde\psi^{\left(1\right)}_{m,2k}$, $\sigma_{m,k}\doteq\tilde\psi^{\left(1\right)}_{m,2k+1}$, $\theta_{m,k}\doteq\tilde\psi^{\left(2\right)}_{m,2k}$, $\chi_{m,k}\doteq\tilde\psi^{\left(2\right)}_{m,2k+1}$,
the Lax pair, choosing the upper sign, can be rewritten as
{\scriptsize\begin{subequations}
\bea
 \nonumber\phi_{m,k+1}&=&{\mathcal L_{m,k}}\phi_{m,k},\ \ \ \phi_{m+1,k}={\mathcal M_{m,k}}\phi_{m,k},\ \ \ \phi_{m,k}\doteq\left(\begin{array}{c}
\rho_{m,k}\\
\theta_{m,k}\end{array}\right),
\\ 
 {\mathcal L_{m,k}}&=&-\alpha_{3}\left(\begin{array}{cc}
1 & w_{m,k+1}-w_{m,k}\\
0 & 1\end{array}\right),\label{Lax4}
\eea
\bea
&&{\mathcal M_{m,k}}=\frac{1}{w_{m,k}-w_{m+1,k}+\ri\epsilon\alpha_{2}}\left(\begin{array}{cc}
\alpha_{3}\left(w_{m,k}-w_{m+1,k}\right)+\alpha_{2}w_{m+1,k} & -\alpha_{2}w_{m,k}w_{m+1,k}-\epsilon^2\alpha_{3}\alpha_{2}\left(\alpha_{3}-\alpha_{2}\right)\\
\alpha_{2} & \alpha_{1}\left(w_{m,k}-w_{m+1,k}\right)-\alpha_{2}w_{m,k}\end{array}\right),\label{Lax5}\\
&& \left(\begin{array}{c}
\sigma_{m,k}\\
\chi_{m,k}\end{array}\right)=\frac{1}{\ri-\epsilon z_{m,k}}\left(\begin{array}{cc}
z_{m,k} & \alpha_{3}-w_{m,k}z_{q,k}\\
1 & -w_{m,k}-\epsilon^2\alpha_{3}z_{m,k}\end{array}\right)\phi_{m,k},\label{Lax6}\\ \\
&&\alpha_{3}{\mathcal E_m}{\mathcal N_{q,k}}\phi_{q,k}=0,\label{Lax7}
\eea
\bea
\nonumber{\mathcal N_{m,k}}\doteq\left(\begin{array}{cc}
\ri\epsilon\left(z_{m,k}-z_{m+1,k}\right) & \ri\epsilon\left(w_{m,k}z_{m+1,k}-w_{m+1,k}z_{m,k}\right)+\epsilon^2\alpha_{3}\left(z_{m,k}-z_{m+1,k}\right)-\epsilon^2\alpha_{2}z_{m,k}+w_{m,k}-w_{m+1,k}+\ri\epsilon\alpha_{2}\\
z_{m,k}-z_{m+1,k} &w_{m+1,k}z_{m+1,k}-w_{m,k}z_{m,k}-\ri\epsilon\alpha_{3}\left(z_{m,k}-z_{m+1,k}\right)-\ri\epsilon\alpha_{2}z_{m+1,k}-\ri\epsilon\left(w_{m,k}-w_{m+1,k}+\ri\epsilon\alpha_{2}\right)z_{m,k}z_{m+1,k}\end{array}\right),
\eea
\end{subequations}}\\
 with $w_{m,k}-w_{m+1,k}+\ri\epsilon\alpha_{2}\not=0$,  $z_{m,k}-z_{m+1,k}\not=0$ and ${\mathcal E_m}$ is the left hand side of the equation (\ref{Krur1}). In deriving (\ref{Lax4}) and (\ref{Lax7}) we have used the relation (\ref{Lax6}) and its difference consequences. The compatibility between (\ref{Lax4}) and (\ref{Lax5}) implies $w_{m,k}-w_{m+1,k}-w_{m,k+1}+w_{m+1,k+1}=0$, while (\ref{Lax7}) implies ${\mathcal E_m}=0$, which is (\ref{Krur1}) %(requiring the matrix multiplying $\phi_{m,k}$ be zero $\forall\alpha_{1}$ turns out to be equivalent asking its determinant be). 
 Analogously one could have written any relation in terms of the vector $\left(\begin{array}{c}
\sigma_{m,k}\\
\chi_{m,k}\end{array}\right)$.  In this case  the compatibility of the so obtained Lax pair, after an integration, would imply  (\ref{Krur3}). 

Let us show that the Lax pair (\ref{Lax4}, \ref{Lax5}) is fake. Under the gauge transformation
\bea
\phi_{m,k}={\mathcal G_{m,k}}\phi^{\prime}_{m,k},\ \ \ {\mathcal G_{m,k}}\doteq\left(-\alpha_{3}^{k}\right)\alpha_{1}^{m}\left(\begin{array}{cc}
1 & w_{m,k}\nonumber\\
0 & 1\end{array}\right),
\eea\\
the Lax pair (\ref{Lax4}, \ref{Lax5}) becomes
\bea
&&\label{Lax4a}{\mathcal L^{\prime}_{m,k}}=\left(\begin{array}{cc}
1 & 0\\
0 & 1\end{array}\right),\\
&&\label{Lax5a}{\mathcal M^{\prime}_{m,k}}=\frac{1}{w_{m,k}-w_{m+1,k}+\ri\epsilon\alpha_{2}}\left(\begin{array}{cc}
w_{m,k}-w_{m+1,k} & \left(w_{m,k}-w_{m+1,k}\right)^2-\epsilon^2\alpha_{2}\left(\alpha_{3}-\alpha_{2}\right)\\
\frac{\alpha_{2}}{\alpha_{3}} & w_{m,k}-w_{m+1,k}\end{array}\right).
\eea\\
In (\ref{Lax4a}, \ref{Lax5a}) ${\mathcal L^{\prime}_{m,k}}$ is independent both on the spectral parameter $\alpha_{3}$ and on any field. So it is a useless  matrix for solving any spectral problem.
%%%%%%%%%%%%%%%%%%%%%%%%%
%%%%%%%%%%%%%%%%%%%%%%%%%
\section{Generalized Symmetries and their B\"acklund transformations.}
%%%%%%%%%%%%%%%%%%%%%%%%%
%%%%%%%%%%%%%%%%%%%%%%%%%
In this Section we construct the three point generalized symmetries written as $$x_{m,n;\lambda}=g_{m,n}\left(s,t,u,v,\bar u,\bar v, x_{m,n};\alpha_{2},\epsilon\right), \qquad s\doteq\frac{x_{m+1,n}-x_{m,n}}{1+\epsilon^2x_{m+1,n}x_{m,n}},$$ $$ t\doteq\frac{x_{m,n}-x_{m-1,n}}{1+\epsilon^2x_{m-1,n}x_{m,n}},\ \ \
 u\doteq x_{m+1,n}-x_{m,n},\ \ \ v\doteq x_{m,n}-x_{m-1,n}$$ $$ \bar u\doteq x_{m,n+1}-x_{m,n},\ \ \ \bar v\doteq x_{m,n}-x_{m,n-1} $$associated to the equation (\ref{eql}) in the $m$ and $n$ direction and then discuss the equation (\ref{eql}) as a B\"acklund transformation for the obtained differential difference equations.$\\$

{\bf Three-points generalized symmetries along direction $m$.}
\begin{itemize}
\item{\bf Case $\epsilon\not=0$:}
{\scriptsize\bea
g_{m,n}^{(\epsilon)}=\Fp{n}\left\{\frac{\alpha_{2}\left(v^2+\epsilon^2\alpha_{2}^2\right)}{\left(u-v\right)\left(u+v\right)}B\left(\frac{\alpha_{2}}{u},m\right)-\frac{\alpha_{2}\left(u^2+\epsilon^2\alpha_{2}^2\right)}{\left(u-v\right)\left(u+v\right)}B\left(\frac{\alpha_{2}}{v},m-1\right)+\right.\label{Sator1}\\
\nonumber\left.+\left[x_{m,n}-\frac{\left(u^2+\epsilon^2\alpha_{2}^2\right)v}{\left(u-v\right)\left(u+v\right)}\right]\omega+\gamma_{n}\right\}+\Fm{n}\left[\frac{s^2t^2}{\left(s-t\right)\left(s+t\right)}\left(B\left(s,m\right)-B\left(t,m-1\right)\right)-\right.\\
\nonumber\left.-\frac{s^2t}{\left(s-t\right)\left(s+t\right)}\omega+\delta_{n}\right]\left(1+\epsilon^2x_{m,n}^2\right), 
\eea}\\
\noindent where $B\left(y,m\right)$, $\gamma_n$ and $\delta_n$ are generic functions of their arguments and $\omega$ is an arbitrary parameter. Let us  note that any function and parameter may eventually depend on $\alpha_{2}$ and $\epsilon$. It is possible to demonstrate that, as long as $\epsilon\not=0$, no $n-$independent reduction of the above symmetry exists when $x_{m,n}\doteq x_m$;\\
\item{\bf Case \; $\epsilon=0$:}

In this case the equations are simpler and we get:
{\scriptsize\bea
g_{m,n}^{(0)}=\frac{u^2v^2}{\left(u-v\right)\left(u+v\right)}\left[\Fp{n}\left(\frac{B\left(\frac{\alpha_{2}}{u},m\right)}{u^2}-\frac{B\left(\frac{\alpha_{2}}{v},m-1\right)}{v^2}\right)\alpha_{2}+\right.\label{Sator2}\\
\nonumber\left.+\Fm{n}\left(B\left(u,m\right)-B\left(v,m-1\right)\right)\right]+\left[\frac{x_{m,n}}{2}-\frac{u^2v}{\left(u-v\right)\left(u+v\right)}\right]\omega+\left(-1\right)^n\sigma x_{m,n}+f_{n},
\eea}\\
\noindent where $B\left(y,m\right)$ and $f_n$ are generic functions of their arguments and $\omega$ and $\sigma$ are arbitrary parameters. As before   any function and parameter may eventually depend on $\alpha_{2}$. It is easy to show that, taking $f_{n}=\frac{1+\left(-1\right)^n}{2}\gamma_{n}+\frac{1-\left(-1\right)^n}{2}\delta_{n}$, the $\epsilon\rightarrow 0$ limit of the symmetry (\ref{Sator1}) coincides with the restriction $\sigma=\omega/2$ of the symmetry (\ref{Sator2}). When $\epsilon=0$ our equation becomes $n-$independent. Hence it is natural to ask if the family of symmetries (\ref{Sator2}) contains a $n-$independent reduction. This reduction exists and is given by
{\scriptsize\bea
g_{m,n}^{(0)}=\frac{u^2v^2}{\left(u-v\right)\left(u+v\right)}\left(\frac{B\left(u,m;\alpha_{2}\right)}{u}-\frac{B\left(v,m-1;\alpha_{2}\right)}{v}\right)+\left[\frac{x_{m,n}}{2}-\frac{u^2v}{\left(u-v\right)\left(u+v\right)}\right]\omega+\chi,\label{Vertumnus},
\eea}\\
\noindent where $\omega$ and $\chi$ are arbitrary parameters (eventually depending on $\alpha_{2}$) and $B\left(y,m;\alpha_{2}\right)$ is a function of its arguments satisfying the following functional equation
\bea
\nonumber B\left(\frac{\alpha_{2}}{y},m;\alpha_{2}\right)=B\left(y,m;\alpha_{2}\right),
\eea
 whose general solution reads
\bea
\nonumber B\left(y,m;\alpha_{2}\right)=\left\{\begin{array}{c}
G\left(y,m;\alpha_{2}\right),\ \ \ y\in{\mathcal B}\left(\alpha_{2}\right)\subset\Omega\left(\alpha_{2}\right),\\
G\left(\frac{\alpha_{2}}{y},m;\alpha_{2}\right),\ \ \ \frac{\alpha_{2}}{y}\in{\mathcal B}\left(\alpha_{2}\right),\ \ \ \ \ \ \ \ \ \end{array}\right. 
\eea
\noindent where $G\left(y,m;\alpha_{2}\right)$ is a generic function of its arguments, $\Omega\left(\alpha_{2}\right)$ represents the following subset of the complex plane
\bea
\nonumber\Omega\left(\alpha_{2}\right)\doteq\left\{z\in{\mathcal C}:\vert z\vert>\sqrt{\vert\alpha_{2}\vert}\cup z=\sqrt{\vert\alpha_{2}\vert}e^{\ri\theta_{z}}:\frac{\theta_{\alpha_{2}}}{2}\leq\theta_{z}\leq\frac{\theta_{\alpha_{2}}}{2}+\pi\right\}
\eea
\noindent and ${\mathcal B}\left(\alpha_{2}\right)$ is a generic subset of $\Omega\left(\alpha_{2}\right)$. In particular, if $B\left(y,m;\alpha_{2}\right)$ is an analytic function in the annulus centered at the origin and of radii $r_{1}\left(\alpha_{2}\right)$ and $r_{2}\left(\alpha_{2}\right)$ so that $r_{1}\left(\alpha_{2}\right)<\sqrt{\vert\alpha_{2}\vert}<r_{2}\left(\alpha_{2}\right)$, it is possible to show that
\bea
\nonumber B\left(y,m;\alpha_{2}\right)=U\left(\frac{\alpha_{2}}{y}+y,m;\alpha_{2}\right),
\eea\\
\noindent $U\left(y,m;\alpha_{2}\right)$ being an analytic function of $y$ in some domain of the complex plane but otherwise generic in all its arguments. For example, if in (\ref{Vertumnus}) we set $\omega=0$ and we choose $U\left(y,m;\alpha_{2}\right)=-1$, we obtain the symmetry
\bea
\nonumber\dot{x}_{m,n}=\frac{\left(x_{m+1,n}-x_{m,n}\right)\left(x_{m,n}-x_{m-1,n}\right)}{x_{m+1,n}-x_{m-1,n}}+\chi, 
\eea
belonging to the list of Volterra-type integrable differential difference equations, cfr. \cite{yamilov2006}, page. 597 (after the translation $x_{m,n}\doteq u_{m,n}+t\chi$). 
\end{itemize}

{\bf Three-points generalized symmetries along direction $n$.}

We have the following symmetries in the $n$ direction:
\begin{equation}
    \begin{aligned}
        g^{(\epsilon)}_{m,n} &=  \Fp{n}\, \left( 
        {B}_{n} \left( \frac { \left( \bar u + \bar v) \right) }{1+{\epsilon}^{2}x_{{m,n+1}}x_{{m,n-1}}}
     \right)+{  \kappa}_{{n}} \right) 
     \\
     &+{  \Fm{n}} \left( 1+
     {\epsilon}^{2}x_{{m,n}}^{2} \right) \left({C}_{n} \left( \bar u+\bar v \right) +{  \lambda}_{{m}} \right),
    \end{aligned}
    \label{eq:symm_m}
\end{equation}
where $B_{n}(y)$ and $C_{n}(y)$ can be arbitrary functions of their argument  and of the
lattice variable $n$. $\kappa_m$ and $\lambda_m$ are arbitrary functions of the lattice variable $m$. In the case $\epsilon=0$ we have:
\bea \label{40}
g_{m,n}^{(0)} = D_n(\bar u + \bar v) + (-1)^n \sigma x_{m,n} + \theta_n, 
\eea
with $D_n(y)$ and $\theta_n$ being arbitrary functions of their argument and $\sigma$ an arbitrary parameter.  It is straightforward to show that, taking $D_n(y)=\Fp{n} B_n(y) + \Fm{n} C_n(y)$ and $\theta_n= \Fp{n} \kappa_n + \Fm{n} \lambda_n$, the $\epsilon \rightarrow 0$ limit of the symmetry  (\ref{eq:symm_m}) coincide with the restriction $\sigma=0$ of the symmetry (\ref{40}).
%%%%%%%%%%%%%%%%%%%%%%%%
\subsection{Differential difference equations and B\"acklund transformations.}
%%%%%%%%%%%%%%%%%%%%%%%%
We now interpret (\ref{Sator1}), when $\epsilon \ne 0$,  as a differential difference equation in $m$. As (\ref{Sator1}) depends also on $n$ we can do so only in the case that the dependence is through $\Fp{n}$ and $\Fm{n}$ as in this case we can overcome it by going over to fields depending on even or odd values of $n$.
In this case we must set $\gamma_n$ and $\delta_n$ to zero and define the new fields $w_m$ and $z_m$ as in Section 3.  So (\ref{Sator1}) becomes the following system of equations:
{\scriptsize\bea
&&w_{m;\lambda}^{(\epsilon)}=\left\{\frac{\alpha_{2}\left(\tilde v^2+\epsilon^2\alpha_{2}^2\right)}{\left(\tilde u-\tilde v\right)\left(\tilde u+\tilde v\right)}B\left(\frac{\alpha_{2}}{\tilde u},m\right)-\frac{\alpha_{2}\left(\tilde u^2+\epsilon^2\alpha_{2}^2\right)}{\left(\tilde u-\tilde v\right)\left(\tilde u+\tilde v\right)}B\left(\frac{\alpha_{2}}{\tilde v},m-1\right)+\right.\label{d1}\\
&&\qquad \qquad \nonumber\left.+\left[w_{m}-\frac{\left(\tilde u^2+\epsilon^2\alpha_{2}^2\right)\tilde v}{\left(\tilde u-\tilde v\right)\left(\tilde u+\tilde v\right)}\right]\alpha\right\}, \\ \label{d2}
&&z_{m;\lambda}^{(\epsilon)}=\left[\frac{\tilde s^2\tilde t^2}{\left(\tilde s-\tilde t\right)\left(\tilde s+\tilde t\right)}\left(B\left(\tilde s,m\right)-B\left(\tilde t,m-1\right)\right)-\frac{\tilde s^2\tilde t}{\left(\tilde s-\tilde t\right)\left(\tilde s+\tilde t\right)}\omega\right]\left(1+\epsilon^2z_{m}^2\right), 
\eea}\\
where
$$ \tilde s\doteq\frac{z_{m+1}-z_{m}}{1+\epsilon^2z_{m+1}z_{m}},\qquad  \tilde t\doteq\frac{z_{m}-z_{m-1}}{1+\epsilon^2z_{m-1}z_{m}},$$ $$
 \tilde u\doteq w_{m+1}-w_{m},\ \ \ \tilde v\doteq w_{m}-w_{m-1}$$
Eqs. (\ref{Krur1}, \ref{Krur2}) turn out to be the B\"acklund transformations of (\ref{d1}, \ref{d2}) relating the solution $(w_m,z_m)$ to a new solution $(\tilde w_m, \tilde z_m)$, i.e.
 \begin{subequations}
        \begin{align}
            \left(w_{m}-w_{m+1}\right)\left(z_{m}-z_{m+1}\right)
            &-\alpha_{2}\epsilon^2z_{m}z_{m+1}-\alpha_{2}=0,
            \label{krur51a}\\
             \left(\tilde w_{m}-\tilde w_{m+1}\right)\left(\tilde z_{m}-\tilde z_{m+1}\right)
            &-\alpha_{2}\epsilon^2\tilde z_{m}\tilde z_{m+1}-\alpha_{2}=0,
            \label{krur51b}\\
            \left(\tilde w_{m}-\tilde w_{m+1}\right)\left(z_{m}-z_{m+1}\right)
            &-\alpha_{2}\epsilon^2z_{m}z_{m+1}-\alpha_{2}=0.
            \label{krur52a}
        \end{align}
        \label{krur5a}
    \end{subequations}
    From (\ref{krur5a}) we evince that the solutions of (\ref{d1}, \ref{d2}) are not independent as are related by (\ref{krur51a}). Then given a solution $(w_m,z_m)$ of the two equations (\ref{d1}, \ref{d2}), solving (\ref{krur52a}) we find $\tilde w_m$ and from (\ref{krur51b}) we find $\tilde z_m$.
   
   A similar result one obtains in the case of (\ref{Sator2}) when $\epsilon=0$. 
   
   In the case of the $n$ symmetries, if $\kappa_m=\lambda_m=0$ the dependence on $m$ is just parametric and thus this phenomena is no more present. We have only one non--autonomous differential difference equation and one non--autonomous B\"acklund transformation
\bea \nonumber
       && w_{n,\lambda} =  \Fp{n}\, 
        {B}_{n} \left( \frac {  w_{n+1} -w_{n-1} }{1+{\epsilon}^{2}w_{n+1}w_{n-1}}
     \right) +  \Fm{n} \left( 1+
     {\epsilon}^{2}w_{n}^{2} \right) C_{n} \left( w_{n+1}-w_{n-1} \right) ,    \\ \nonumber
&&\left(w_{n}-\tilde w_{n}\right)\left(w_{n+1}-\tilde w_{n+1}\right)-\epsilon^2\alpha_{2}\left({\mathcal F}_{n}^{\left(+\right)}w_{n+1}\tilde w_{n+1}+{\mathcal F}_{n}^{\left(-\right)}w_{n}\tilde w_{n}\right)-\alpha_{2}=0.
\eea

%%%%%%%%%%%%%%%%%%%%%%%%%
%%%%%%%%%%%%%%%%%%%%%%%%%
\section{Conclusions.}
%%%%%%%%%%%%%%%%%%%%%%%%%
%%%%%%%%%%%%%%%%%%%%%%%%%
In this article we presented results on the linearization and on the symmetries for a non--autonomous nonlinear partial difference equation belonging to the Boll classification of quad--graph equations on the lattice, the ${}_{t}H_{1}^\epsilon$ equation. 

In particular we show its explicit linearization obtained by reducing the ${}_{t}H_{1}^\epsilon$ equation to a system of autonomous partial difference equations which can be explicitly solved and by showing that its Lax pair is fake i.e. by a gauge transformation the spectral problems turns out to be independent from the spectral parameter and the field.

We can find its three point generalized symmetries which reflect the linearizability of the discrete equation by depending on arbitrary functions of a continuous variable and on the discrete lattice index. We plan to discuss in detail this very interesting peculiar result in a future work.

Left to future work is also the analysis of the other linearizable equations of the Boll classification. 
\subsection*{Acknowledgments}
DL  and CS have been partly supported by the Italian Ministry of Education and Research, 2010 PRIN {\it Continuous and discrete nonlinear integrable evolutions: from water waves to symplectic maps}. GG and DL are also supported by INFN IS-CSN4 {\it Mathematical Methods of Nonlinear Physics}.

%%%%%%%%%%%%%%%%%%%%%%%%%%%%%%%

\


\begin{thebibliography}{99}
%%%%%%%%%%%%%%%%%%%%%%%%%%%%%%%
\bibitem{ABS03} V.E. Adler, A.I. Bobenko, Yu.B. Suris \emph{Classification of integrable equations on quad-graphs. The consistency approach}, Comm. Math. Phys. {\bf 233}, 513–543 (2003).
\bibitem{ABS09} V. E. Adler, A. I. Bobenko, Yu. B. Suris \emph{Discrete nonlinear hyperbolic equations. Classification of integrable cases}, Funct. Anal. Appl. {\bf 43}, 3–17 (2009).
 
\bibitem{ABS2009} Adler, V. E. and Bobenko, A. I. and Suris, Y. B. , Discrete nonlinear hyperbolic equations. Classification of integrable cases, {\it Funct. Anal. Apll.}  {\bf 43} (2009) 3--17.

\bibitem{ABS2011}Adler, V. E. and Bobenko, A. I. and Suris, Y. B. ,
Classification of the integrable discrete equations of the octahedron type, {\it Intern. Math. Research Notices} {\bf 60} (2011) 363--401.

\bibitem{At09} J. Atkinson, \emph{Linear quadrilateral lattice equations and multidimensional consistency}, J. Phys. A: Math. Theor. {\bf 42}, 454005 (2009).

\bibitem{Bobenko2008book}Bobenko, A. I. and Schr\"oder, P. and Sullivan, J. M. and Ziegler, G. M., {\it Discrete differential geometry}, Birkh\"auser Verlag AG 2008.

\bibitem{BoS} A.I. Bobenko, Yu.B. Suris, \emph{Integrable systems on quad-graphs}, Int. Math. Res. Notices {\bf 11}, 573–611 (2002).

\bibitem{Boll11} R. Boll, \emph{Classification of $3D$ consistent quad-equations}, J. Nonl. Math. Phys. {\bf 18}, no. 3, 337-365 (2011).
\bibitem{Boll12a} R. Boll, \emph{Corrigendum Classification of $3D$ consistent quad-equations}, J. Nonl. Math. Phys. {\bf 19}, no. 4, 1292001 (2012).
\bibitem{Boll12b} R. Boll, \emph{Classification and Lagrangian structure od $3D$ consistent quad-equations}, Ph. D. dissertation (2012).
\bibitem{BrH} T. Bridgman, W. Hereman, G. R. W. Quispel, P. H. van der Kamp, \emph{Symbolic computation of Lax pairs of partial difference equations using consistency around the cube}, Found. Comput. Math. {\bf 13}, no. 4, 517-544 (2013).

\bibitem{cn91}
F. Calogero and M.C. Nucci,
\newblock Lax pairs galore,
\newblock {\em J. Math. Phys.}, {\bf 32}  (1991) 72-74.

\bibitem{ds}A. Doliwa, P.M. Santini, Multidimensional quadrilateral lattices are
integrable, {\it Phys. Lett. A} {\bf 233}, 365--372 (1997).

\bibitem{gubbiotti_thesis} G. Gubbiotti PhD dissertation to be submitted  in 2017.

\bibitem{GubHay}Gubbiotti, G. and Hay, M., A {\it SymPy}~module to calculate algebraic entropy for difference 
       equations and quadrilateral partial difference equations, in preparation.

\bibitem{GLS2015} G. Gubbiotti, C. Scimiterna and D. Levi, On  quad equations  consistent on the cube, submitted to {\it J. Phys. A: Math. Gen.} 2015.

\bibitem{hb} M. Hay and  S. Butler, Simple identification of fake Lax pair, arXiv:1311.2406v1.

\bibitem{H04} J. Hietarinta, \emph{A new two-dimensional lattice model that is \textquoteleft consistent around the cube\textquoteright}, J. Phys. A: Math. Gen. {\bf 37}, L67-L73 (2004).
\bibitem{H05} J. Hietarinta, \emph{Searching for CAC-maps}, J. Nonl. Math. Phys. {\bf 12}, no. 2, 223-230 (2005).

\bibitem{HietarintaViallet2007}Hietarinta, J. and Viallet, C.,
Searching for integrable lattice maps using factorization, {\it J. Phys. A: Math. Theor.} {\bf 40} (2007) 12629--12643.

\bibitem{HV} J. Hietarinta, C. Viallet, \emph{Weak Lax pairs for lattice equations}, Nonlin. {\bf 25}, 1955-1966 (2012).

\bibitem{lb}D Levi, R Benguria, {\it B\"acklund transformations and nonlinear differential difference equations} Proceedings of the National Academy of Sciences 77 (9), 5025-5027

\bibitem{lsz90}
D. Levi, A. Sym and G.Z. Tu,
\newblock A working algorithm to isolate integrable surfaces in E3, 
\newblock preprint DF INFN 761, Roma, Oct. 10, 1990.

\bibitem{lll10}
Y.Q. Li, B. Li and S.Y. Lou,
\newblock Constraints for Evolution Equations with Some Special Forms of Lax Pairs and Distinguishing Lax Pairs by Available Constraints,
\newblock Eprint (2010) arXiv:1008.1375v2 [nlin.SI].

\bibitem{m10}
M. Marvan,
\newblock On the spectral parameter problem,
\newblock {\em Acta Appl Math} {\bf 109} (2010) 239--255

\bibitem{m04}
M. Marvan,
\newblock Reducibility of zero curvature representations with application to recursion operators,
\newblock {\em Acta Appl. Math.},  {\bf 83} (2004) 39--68.


\bibitem{N} F.W. Nijhoff, \emph{Lax pair for the Adler (lattice Krichever- Novikov) system}, Phys. Lett. A {\bf 297}, 49–58 (2002).
\bibitem{NW} F.W. Nijhoff, A. Walker, \emph{The discrete and continuous Painlev\'e VI hierarchy and the Garnier systems}, Glasgow Math. J. 43A, 109 (2001).

\bibitem{ns}JJC Nimmo, WK Schief, Superposition principles associated with the Moutard transformation: an integrable discretization of a (2+ 1)--dimensional sine--Gordon system, Proceedings of the Royal Society of London A: Mathematical, Physical and Engineering Sciences, {\bf 453} (1997) Pages
255-279.


\bibitem{RJ} A. Ramani, N. Joshi, B. Grammaticos, T. Tamizhmani, \emph{Deconstructing an integrable lattice equation}, J. Phys. A: Math. Gen. {\bf 39}, L145-L149 (2006).

\bibitem{s01}
S.Yu. Sakovich,
\newblock True and fake Lax pairs: how to distinguish them, 
\newblock arXiv:nlin.SI/0112027.

\bibitem{s02}
S.Yu. Sakovich,
\newblock Cyclic bases of zero-curvature representations: five illustrations to one concept,
\newblock arXiv:nlin/0212019v1.


\bibitem{Tremblay2001}Tremblay, S. and Grammaticos, B. and Ramani, A., Integrable lattice equations and their growth properties, {\it Phys. Lett. A} {\bf 278} (2001) 319--324.

\bibitem{Viallet2006}Viallet, C.,  Algebraic  Entropy for lattice equations, \texttt{arXiv:math-ph/0609043}.

\bibitem{XP} P. D. Xenitidis, V. G. Papageorgiou, \emph{Symmetries and integrability of discrete equations defined on a black-white lattice}, J. Phys. A: Math. Theor. {\bf 42}, 454025 (2009).

\bibitem{yamilov2006} R. Yamilov, \emph{Symmetries as integrability criteria for differential difference equations}, J. Phys. A: Math. Gen. {\bf 39}, R541-R623 (2006)
%\bibitem{Boll2011} Boll, R. , Classification of {3D} consistent quad-equations, {\it J. Nonlinear Math. Phys.} {\bf 18} (2011) 337--365}.

%\bibitem{Boll-th}Boll, R. ,{\it Classification and Lagrangian Structure of 3D Consistent Quad-Equations} PhD Thesis Technische Universität Berlin 2012.

%\bibitem{Bridgman2013} {Bridgman, T. and Hereman, W. and Quispel, G.R.W. and van der Kamp, P.H., Symbolic Computation of Lax Pairs of Partial Difference Equations using Consistency Around the Cube, {\it Foundations of Computational Mathematics} {\bf 13} (2013) 1615--3375.

%\bibitem{HietarintaViallet2012}Hietarinta, J. and Viallet, C., Weak {\it Lax} pairs for lattice equations,  {\it Nonlinearity} {\bf 25} (2012) 1955--1966}.

%\bibitem{Nijhoff2001}Nijohf, F. W. and Walker, A. J.,The discrete and continous {Painlev\`e} hierarchy and the {Garnier} systems., {\it Glasg. Math. J.} {\bf 43A} (2001) 109--123.



%\bibitem{Xenitidis2009}Xenitidis, P. D. and Papageorgiou, V. G.,Symmetries and integrability of discrete equations defined on a black–white lattice,  {\it J. Phys. A: Math. Theor.} {\bf 42} (2009) 454025}.


\end{thebibliography}
\end{document}